\documentclass[useAMS,usenatbib]{mn2e}
\usepackage{graphicx}
\usepackage{natbib}
\usepackage{ amssymb}
\usepackage{epstopdf}
\title[The End of an Era]{The End of an Era - The Population III to Population II Transition and the Near Infrared Background}
\author[Elizabeth R. Fernandez and Saleem Zaroubi]{Elizabeth R. Fernandez\thanks{E.Fernandez@astro.rug.nl} and Saleem Zaroubi\\
Kapteyn Astronomical Institute, The University of Groningen,  
Landleven 12,               
9747 AD Groningen,       
The Netherlands
}%
\begin{document}
\label{firstpage}
\maketitle
\begin{abstract}
There are only a few ways to constrain the Era of Reionization and the properties of high redshift ($z \gtrsim 6$) stars through observations.  Here, we discuss one of these observables - the spectrum of the Near Infrared Background - and how it is potentially affected by the transition from Population III to Population II stars.  The stronger Lyman-$\alpha$ emission expected from massive Population III stars could result in a 'bump' in the spectrum of the Near Infrared Background (referred to in this work as the Lyman-$\alpha$ bump).  The strength and shape of this bump can reveal properties of Population III stars.  The Lyman-$\alpha$ bump is predicted to be higher if Population III stars are more massive and present at lower redshifts.  The shape of the bump is governed by the star formation rate and the time it takes Population III stars to transition to Population II stars.  If Population III stars are indeed  massive, a bump is predicted as long as Population III stars exist at $z \lesssim 15$, even if their star formation rate is as low as $10^{-7} \; \rm{M_\odot yr^{-1} Mpc^{-3}}$.  This means that there may be some observational signature in the Near Infrared Background of small pockets of metal-free gas forming Population III stars at $z \sim 6$, even if they are quite rare.  
\end{abstract}
\section{Introduction}
\label{sec:introduction}
The first generation of stars, known as Population III, were truly metal free.  These stars  have never been observed, so their properties must be surmised using indirect or theoretical means.  It is possible that these true first generation of stars were of the order of hundreds of solar masses \citep{abel/etal:2002, bromm/etal:2002, glover:2005}, although some suggest these stars were less massive with a broader initial mass function (IMF; \citet{clark/etal:2008, clark/etal:2011, stacy/etal:2010, greif/etal:2011, hosokawa/etal:2011}).  
As star formation progressed and these stars died, the intergalactic medium was enriched with metals.  After several generations of enrichment, the amount of metals present would be significant enough to act as an efficient coolant, causing the properties of subsequent generations of stars to be fundamentally different.  These Population II stars, forming in clouds that can undergo metal and dust cooling, were able to fragment into smaller masses, likely causing the masses of these stars to be less than their Population III progenitors \citep{bromm/etal:2002,schneider/etal:2002, bromm/loeb:2003}.  Currently, it is unknown when this transition from Population III to Population II stars occurred and how extended in time it was.  However, even though the majority of star formation transitioned to Population II, it is possible that Population III formation persisted even to $z \sim 6$ in pockets of pristine gas \citep{tornatore/etal:2007, trenti/etal:2009}.  

Observing this era of high-redshift star formation is very challenging.  High redshift galaxy surveys are now routinely discovering galaxies at very high redshifts ($z>6$).  However, these detections are currently limited to only the brightest objects common enough to appear in the survey field, missing out on the bulk of numerous, smaller galaxies.  These galaxies below the detection limit could be responsible for a significant amount of star formation, and could be the primary drivers of reionization \citep{barkana/loeb:2000, salvaterra/ferrara:2006, wyithe/loeb:2006, kistler/etal:2009, Bouwens/etal:2010,Robertson/etal:2010,fernandez/shull:2011,munoz/loeb:2011}.  Even though these galaxies are below the limiting magnitude of current surveys, it may be possible to observe their redshifted cumulative spectrum, which would be present in any background emission in the infrared.  Therefore, understanding what portion of the Near Infrared Background (NIRB) is due to a high redshift component could lead to important constraints on these early stars and galaxies.  

Actually measuring the contribution of these high redshift stars to the NIRB is no easy task, due to great uncertainties in foreground subtraction.  In order to accurately establish which portion of the spectrum is actually due to stars during the era of reionization, low redshift galaxies must be correctly subtracted.  Zodiacal light, which is very difficult to model, must also be taken into account.   
Yet, despite the difficulties, many have undertaken the challenging observation to measure this excess of the NIRB that is not attributed to lower redshift objects ($z\lesssim 6$) and other foregrounds \citep{dwek/arendt:1998,gorjian/wright/chary:2000,kashlinsky/odenwald:2000,wright/reese:2000,
wright:2001,cambresy/etal:2001,totani/etal:2001, kashlinsky:2005,  magliocchetti/salvaterra/ferrara:2003, odenwald/etal:2003,cooray/etal:2004, matsumoto/etal:2005, kashlinsky/etal:2002,kashlinsky/etal:2004, kash/etal:2007, kashlinskyb/etal:2007c, kashlinsky/etal:2012}.  These observations remain controversial, with some claiming that the NIRB is resolved - either in lower redshift galaxies \citep{thompson/etal:2007a,thompson/etal:2007b} or intrahalo stars \citep{cooray/etal:2012a}.

However, because ultraviolet photons are needed to reionize the Universe, any signal from the Epoch of Reionization would be redshifted and present in the NIRB, even if it is not detectable with current instruments.  In this spirit, many have sought to model the contribution of stars and galaxies responsible for reionization on the NIRB \citep{santos/bromm/kamionkowski:2002, magliocchetti/salvaterra/ferrara:2003,salvaterra/ferrara:2003,
cooray/etal:2004,cooray/yoshida:2004,kashlinsky:2005,madau/silk:2005, fernandez/komatsu:2006,  thompson/etal:2007a,thompson/etal:2007b, fernandez/etal:2010, fernandez/etal:2012fluc, fernandez/etal:2012, cooray/etal:2012, kashlinsky/etal:2002, kashlinsky/etal:2004,kashlinsky/etal:2005,kashlinsky/etal:2007,kashlinsky/etal:2012,yue/etal:2012}.
New observations will better constrain any observable contribution of stars from the Epoch of Reionization to the NIRB, such as from the Cosmic Infrared Background ExpeRiment (CIBER), AKARI, and the Cosmic Assembly Near-infrared Deep Extragalactic Legacy Survey (CANDLES).  

In this work, we explore how the transition from Population III to Population II stars is imprinted on the spectrum of the NIRB.  Our stellar models, in addition to describing the star formation rate and how we deal with the transition from Population III to Population II stars, are described in \textsection \ref{sec:models}.  Our formalism for obtaining the spectrum of the NIRB is detailed in \textsection \ref{sec:spectra}.  The effect of various parameters on the NIRB spectrum, such as the mass of stars, the dependence the star formation rate, the contribution of the intergalactic medium (IGM), and parameters of the Population III to Population II transition, are detailed in \textsection \ref{sec:results}.  We conclude in \textsection \ref{sec:conclusions}.

\section{Stellar Population Models}
\label{sec:models}
In order to understand the spectra of high redshift stars, we need to establish stellar properties, such as mass and metallicity.  This metallicity will increase as time goes on, so a model for the transition from metal-free to metal-poor stars must be established.  We also must model how many stars are forming and how this changes with redshift.

\subsection{Mass and Metallicity of High Redshift Stars}
\label{sec:mass}

In order to model the various stellar populations,  
we establish some limiting cases for both the mass and metallicity of the stars.  For the metallicity of the stars, we use either Population III stars (with no metals), or Population II metal-poor stars (with $Z=1/50 \; \rm{Z_\odot}$), as in \citet{fernandez/komatsu:2006}.  
Since the mass of Population III stars is still largely unknown, we use three possibilities for the initial mass spectra.  The first two are represented by a Larson mass spectrum \citep{larson:1998}:
\begin{equation}
f(m)\propto m^{-1}\left(1+\frac{m}{m_c}\right)^{-1.35}.
\label{eq:larson}
\end{equation}
For a massive population, we set the mass limits to $m_1 = 0.1 \; \rm{M_\odot}$, $m_2=500\; \rm{M_\odot}$, and $m_c = 250\; \rm{M_\odot}$, while for a less massive population, we use $m_1 = 0.1 \; \rm{M_\odot}$, $m_2=150\; \rm{M_\odot}$, and $m_c = 10\; \rm{M_\odot}$.  For the case with the least massive Population III stars, along with Population II stars, we use a Salpeter mass function \citep{salpeter:1955}:
\begin{equation}
f(m) \propto m^{-2.35},
\label{eq:salpeter}
\end{equation}
with mass limits of $m_1=0.1 \; \rm{M_\odot}$ and $m_2=150 \; \rm{M_\odot}$. \footnote{Changing the lower mass limit, while not changing the overall number of photons produced, will change the star formation rate, since more mass will be tied up in low mass stars.  Therefore, a higher value of $m_1$ will cause a lower overall star formation rate for a given number of ionizing photons produced.  Since all the ionizing photons are from stars above $5 M_\odot$, our choice of $m_1$ only has a small effect on the overall spectrum.}  

\subsection{The Star Formation Rate}

The star formation rate at high redshifts is very hard to probe observationally.  There are some results deduced from high redshift galaxy surveys \citep[e.g.][]{hopkins/beacom:2006, mannucci/etal:2007, bouwens/etal:2008,Bunker/etal:2010,bouwens/etal:2011,bouwens/etal:2011b,bouwens/etal:2012,mclure/etal:2012, zheng/etal:2012, Finkelstein/etal:2012, oesch/etal:2012,oesch/etal:2013, ellis/etal:2013} up to a redshift of $z\sim 12$.  However, these results may not probe the population as a whole (including the smallest of galaxies).  Higher rates are deduced from gamma-ray bursts \citep[e.g.][]{kistler/etal:2009, robertson/ellis:2012}, which can also infer the star formation rate in fainter galaxies \citep[e.g.][]{kistler/etal:2009, trenti/etal:2012}.  The star formation rate at even higher redshifts is still unconstrained. 

\begin{figure}
\centering \noindent
\includegraphics[width=8.8cm]{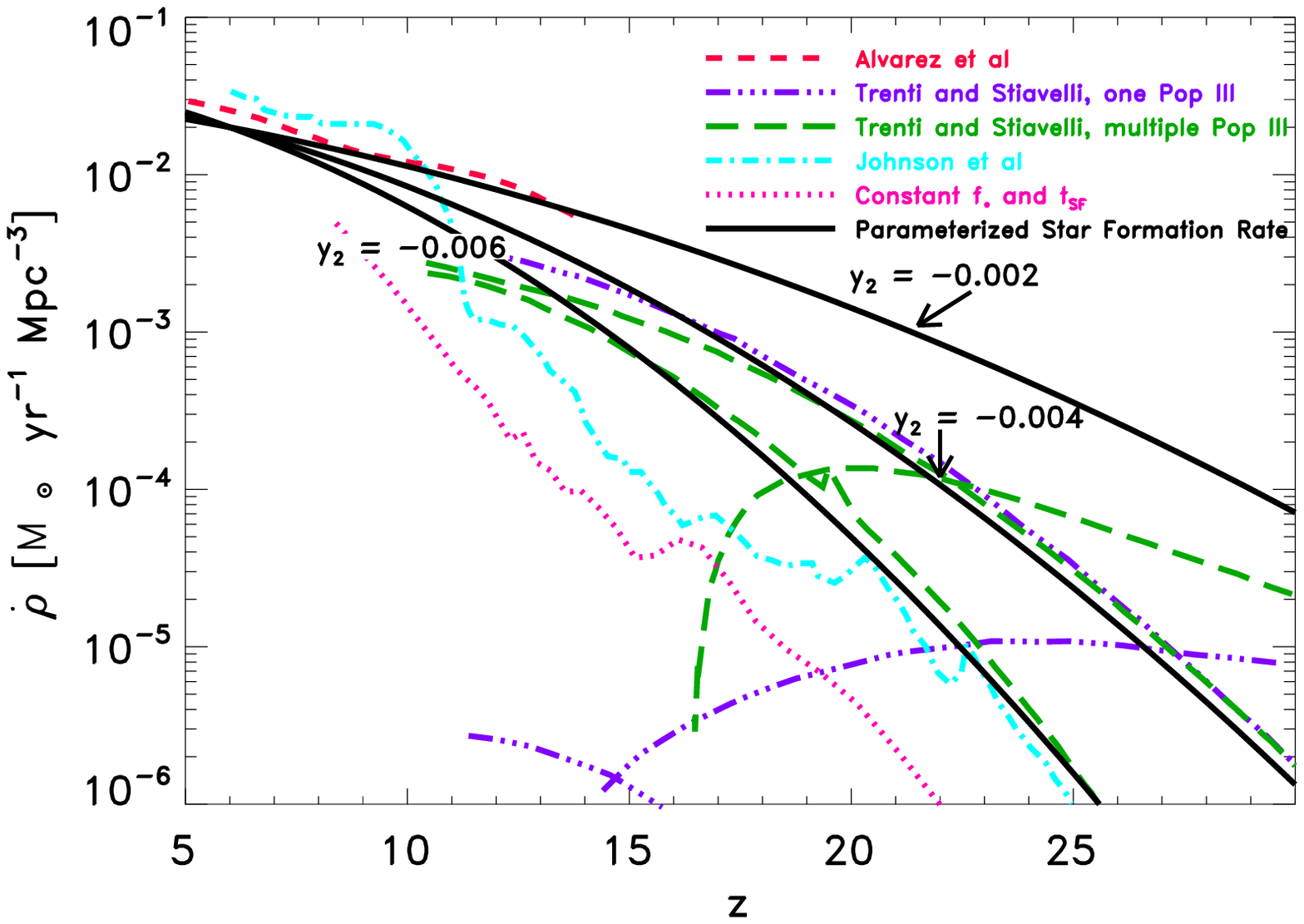}
\caption{Our star formation rate models, compared with results from analytical models \citep{trenti/stiavelli:2009, alvarez/etal:2012} and simulations  \citep{johnson/etal:2013}.  The parametrized models are the models we use, based on equation \ref{eq:sfrapp}, which are shown as the solid lines for  $y_2 = -0.002$, $-0.004$, or $-0.006$.  The pink dotted line shows our second model for the star formation rate when a constant value of $f_*=0.003$ and a star formation time-scale of $11.5 \; \rm{Myr}$ is assumed. }
\label{fig:varsfra}
\end{figure}

Further knowledge of the star formation rate is advanced through theoretical means.  In Fig. \ref{fig:varsfra}, we show the star formation rate based on analytical models from \citet{trenti/stiavelli:2009}, who modelled star formation based on the collapse fraction of haloes, along with gas cooling, radiative feedback, and metal enrichment.  The data was taken from their standard model, which contains only one Population III star per halo (purple dashed-triple dotted lines), or multiple Population III stars per halo (green long dashed lines).  They calculated the values of the star formation rate of Population II stars (top lines for each colour) and Population III stars in either mini-haloes or more massive haloes with $T_{vir} \geq 10^4K$ (the lower two lines of each colour).  The cyan dot dashed line \citep{johnson/etal:2013} shows the result of cosmological simulations that contain metal enrichment and stellar radiation fields.  The model from  \citep{alvarez/etal:2012} (the red short dashed line) shows the star formation rate for Population II stars only from models which take into account suppression on small galaxies within ionized regions.  

Then, in order to model the star formation rate $\dot{\rho_*}(z)$, we approximate a fit as:
\begin{equation}
\dot{\rho_*}(z) = 10^{y_0 + y_1 z + y_2 z^2},
\label{eq:sfrapp}
\end{equation}
where $y_0$, $y_1$, and $y_2$ are constants.  This model is not meant to be physical, but instead it approximates the star formation rate to be consistent with results from these analytical formulations and simulations.  We set $y_1 = -0.03$, and change $y_2$ to be either $-0.002$, $-0.004$, or $-0.006$.  This effectively changes the steepness of the slope of the star formation rate, where more negative values of $y_2$ lead to less star formation at high redshift.  The value of $y_0$ is then set so that the star formation rate at $z=6$ is $0.02$ $\; \rm{M_\odot yr^{-1} Mpc^{-3}}$, approximately consistent with results from theory \citep{alvarez/etal:2012, johnson/etal:2013}, and slightly higher than current observational constraints from gamma-ray bursts \citep[e.g.][]{kistler/etal:2009, robertson/ellis:2012}.  These models are shown as the solid black lines, also in Fig.  \ref{fig:varsfra}.  Finally, as another model for the star formation rate, we also assume a constant star formation efficiency over redshift, which has often been assumed in the literature.  This is shown as the dotted pink line.  (This model will be discussed more in Section \ref{sec:move}).     

\subsection{The Transition from Population III to Population II Stars}

\begin{figure}
\centering \noindent
\includegraphics[width=8.8cm]{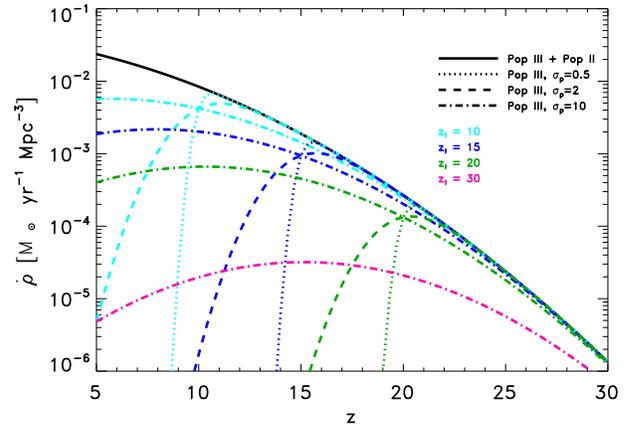}
\caption{ The total star formation rate of Population III and Population II stars with $y_2 = -0.004$ (solid black line) and Population III stars only (coloured lines) for various values of the redshift of transition from Population III to Population II stars ($z_t$) and the transition width ($\sigma_p$).  }
\label{fig:varsfrb}
\end{figure}

We assume that the transition of Population III to Population II stars is described by the formula:
\begin{equation}
f_p = \frac{1}{2} [ 1 + erf(\frac{z-z_t}{\sigma_p} )]
\label{eq:fp}
\end{equation}
\citep{cooray/etal:2012} where $f_p$ is the fraction of stars that are Population III.  This formula simplifies the likely very in-homogeneous transition with only a few free parameters which describe the redshift where Population III stars transitioned to Population II stars ($z_t$) and the length of this transition (parametrized by $\sigma_t$).  We then define the star formation rate of Population III stars as $f_p \dot{\rho}(z)$ and the star formation rate of Population II stars as $(1-f_p)\dot{\rho}(z)$.  Both $z_t$ and $\sigma_t$ can be adjusted to simulate various enrichment histories.  
We show our total star formation rate (given by the solid black line in Fig. \ref{fig:varsfrb}) for $y_2 = -0.0004$, along with the star formation rate of Population III stars only (given as the coloured lines) representing various values of $z_t$ and $\sigma_p$.

\subsection{Constraints from Reionization}
Properties of the stars themselves, such as the number of ionizing photons produced per second per star, $Q(H)$, and the lifetime of the star, $\tau_*$, are taken from stellar models from \citet{schaerer:2002}, or from the fitting formulas in \citet{fernandez/komatsu:2008} based on the stellar models in \citet{lejeune/schaerer:2001}. When combining these models with the mass spectrum $f(m)$ defined in section \ref{sec:mass}, we can compute the rate of ionizing photons produced per comoving volume ($\Gamma$).  The formula used depends on the mass of the star.  Stars that are massive (above a mass $m_t(z)$, whose lifetimes are shorter than the star formation time-scale, here assumed to be the age of the Universe) have a photon production rate based on the star formation rate $\dot{\rho_*}(z)$:
\begin{equation}
\Gamma_l = \frac{\int^{m_2}_{m_t(z)} Q(H) \tau_*(m) f(m) dm}{ \int_{m_1}^{m_2}dm f(m)m}  \dot{\rho}_*(z)
\end{equation}
Less massive stars (denoted by the subscript $s$), whose lifetimes are longer than the age of the Universe, do not emit all of their ionizing photons before the end of reionization.  Therefore, the number of ionizing photons per stellar mass is computed instead by using the comoving mass density of stars ($\rho_*(z)$):
\begin{equation}
\Gamma_s = \frac{\int^{m_t}_{m_1} Q(H) f(m) dm}{ \int_{m_1}^{m_2}dm f(m)m} \rho_*(z)
\end{equation}
where $\rho_*(z)=\int^{z_2}_z \dot{\rho_*}(z)/[H(z)(1+z)]dz$.  We assume that the ratio of ionizing photons that escape into the IGM by the end of reionization to baryons ($\Upsilon$) is 2, which is known as photon-starved reionization \citep{Bolton/Haehnelt:2007}, derived from a combination of simulations, measurements of the Lyman-$\alpha$ effective optical depth, and an assumption of the mean free path.  We can then solve for the escape fraction over the interval from $z_2$, where star formation starts, to $z_1$, where reionization is complete: 
\begin{equation}
f_{esc} = \frac{\Upsilon n_H}{ \int_{z_1}^{z_2} dz [\Gamma_{III}(z) + \Gamma_{II}(z)] / [(1+z)H(z)]}.
\end{equation}
For the purposes of this paper, we take $z_2$ to be 30, and $z_1$ to be 6.  Therefore, the number density of hydrogen ($n_H$) is balanced by the number of photons produced by each population of stars, where $\Gamma_{III}$ or $\Gamma_{II}$ is the sum of $\Gamma_l$ and $\Gamma_s$ for either Population III or Population II stars.  

\section{The Intensity of the Near Infrared Background}
\label{sec:spectra}

Now that our stellar models are established, we can compute the portion of the NIRB expected due to high redshift stars within our redshift range ($6<z<30$).  This emission would result from a combination of various stellar and nebular processes.  
The intensity of the NIRB is given as:
\begin{equation}
 I_{\nu} =
 \frac{c}{4\pi} 
 \int 
 \frac{dz\, p([1+z]\nu, z)}{H(z) (1+z)},
\label{eq-inu-generic}
\end{equation}
\citep{peacock:1999}.  In order to compute the component of the intensity of the NIRB from high redshift stars, we follow the formalism presented in \citet{fernandez/komatsu:2006}.  Again, in order to compute the emissivity $p([1+z]\nu, z)$,  we must divide our calculation into parts, depending on the lifetime of the star.  If the lifetime of the star is shorter than the star formation time-scale, the emissivity is given by the luminosity $L^\alpha(m)$ of each stellar or nebular component $\alpha$, integrated over a mass spectrum of stars $f(m)$, and is weighted by the lifetime of the star $\tau_*(m)$.  This is normalized by $m_*$, or the mean stellar mass of the mass spectrum:
\begin{equation}
  p(\nu,z)  
= \dot{\rho}_*(z)c^2 \frac1{m_*}\int dm~mf(m)
  \left[\frac{\overline{L}^\alpha_\nu(m)\tau_*(m)}{mc^2}\right],
\end{equation}
and if the lifetime of the star is longer than the star formation time-scale, the emissivity is given by:
\begin{equation}
  p(\nu,z)  
=
 \rho_*(z)c^2 \frac{1}{m_*}\int dm~mf(m)
  \left[\frac{\overline{L}^\alpha_\nu(m)}{mc^2}\right].
\end{equation}
(See appendix A of \citet{fernandez/komatsu:2006}).  
Emission will originate from the star itself (in the form of a stellar blackbody), as well as reprocessed nebular emission, namely, free-free, free-bound, and two-photon emission, along with emission lines, the primary one being the Lyman-alpha line.  The luminosity of each component $\overline{L}^\alpha_\nu(m)$ is modelled as in \citet{fernandez/komatsu:2006}.  For our initial cases, we assume that only emission from the haloes contribute to the spectrum of the NIRB, so the luminosity of the nebular components will be multiplied by the factor of $(1-f_{esc})$.

\section{The Spectral Features of the NIRB}
 \label{sec:results}

For our reference model, we allow star formation to begin with only Population III stars with a Larson mass spectrum and $m_c$ of $250$ (see equation \ref{eq:larson}).  At $z_t=10$, star formation transitions to Population II stars via equation \ref{eq:fp}, with a transition width of $\sigma_p=2$.  The star formation rate is assumed to follow equation \ref{eq:sfrapp} with $y_2=-0.004$, and $f_{esc}$ is set to be consistent with photon-starved reionization (in this case, $f_{esc} = 0.16$).  These assumptions (the masses of the stars, details of the transition from Population III to Population II stars, and the star formation rate) will be adjusted in the following sections.  

\subsection{Moving from Population III to Population II Star Formation}
\label{sec:move}

\begin{figure*}
\centering \noindent
\includegraphics{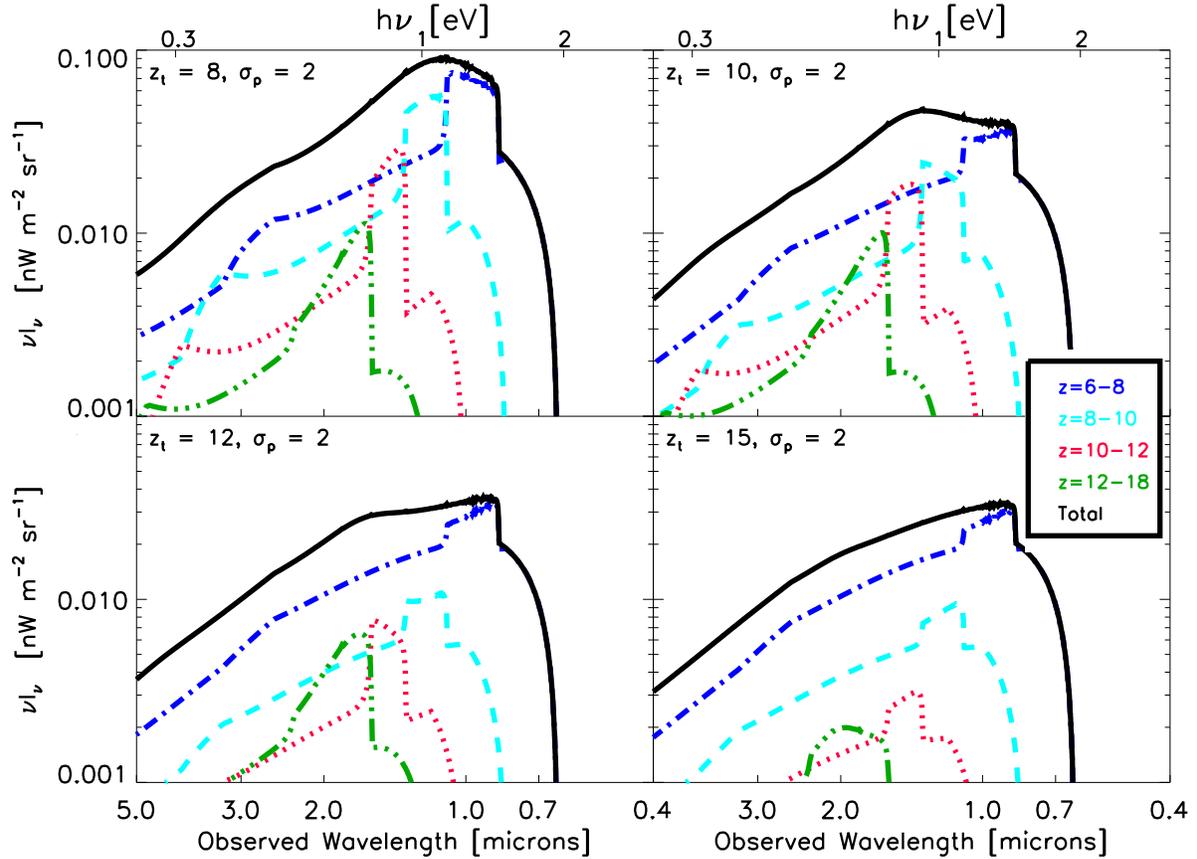}
\caption{The spectrum of the NIRB resulting from stars above a redshift of $6$.  Various cases are shown illustrating different times when the majority of star formation moves from Population III to Population II.  This is quantified by the redshift of transition $z_t$, which describes the fraction of Population III stars in equation \ref{eq:fp}.  If $z_t$ is relatively low, a 'bump' in the spectrum is seen from the end of the Population III epoch.  The solid lines are the total contribution from $z=6-30$, dot-dashed are $z=6-8$, dashed are $z=8-10$, dotted are $z=10-12$, and dashed-triple dot are $z=12-18$.}
\label{fig:zt}
\end{figure*}

It is unknown when the majority of star formation transitioned from Population III to Population II stars.  In Fig. \ref{fig:zt}, we allow this redshift of transition ($z_t$) to be either $8$, $10$, $12$, or $15$.  For our reference case, Population III stars are very massive and will have very strong nebular features, such as the Lyman-$\alpha$ line.  Because of this, it is possible that the overall spectral intensity for a more distant population of Population III stars could have a higher intensity than closer, Population II stars.  
Therefore, the spectrum of the NIRB could contain a 'bump' from the strong Lyman-$\alpha$ line resulting from Population III stars, even though these are at higher redshift than closer, Population II stars.  (A similar prominent Lyman-$\alpha$ feature is seen by \citet{dwek/etal:2005}, who assume all stars forming at high redshift are Population III stars.)   
It is also important to notice that various redshift intervals contribute at different wavelengths.  Especially when $z_t \lesssim 10$, we see that the contribution from stars at larger redshifts ($z>8$) is actually greater than that of stars at $z<8$ at wavelengths of longer than about 1 $\mu$m.  If $z_t$ is larger, the 'Lyman-$\alpha$' bump is less pronounced since these Population III stars are more distant, and the overall spectrum of the NIRB becomes increasingly dominated by Population II stars with a less prominent Lyman-$\alpha$ line.  In addition, as $z_t$ increases, not only does the Lyman-$\alpha$ bump become less pronounced, but the wavelength at which the bump is located increases, corresponding to more distant Population III stars.  As $z_t$ rises, the spectrum of the NIRB becomes increasingly featureless at longer wavelengths.  If $z_t \gtrsim 15$, no Lyman-$\alpha$ bump is seen for our given star formation rate and initial mass spectrum.  

This has interesting observational consequences.  If there are some Population III stars at relatively low redshift, where $z_t \lesssim 10$, they could contribute a signature to the NIRB spectrum, causing it to deviate from a featureless spectrum.  This occurs even if the star formation rate of these stars is quite low.  For example, our reference case has $\dot{\rho_*}_{III}(z=6) \sim 10^{-5} \; \rm{M_\odot yr^{-1} Mpc^{-3}}$, (in other words, the Population III star formation rate is 0.2 per cent of the total star formation rate at $z=6$).  Our case where $z_t=12$ has $\dot{\rho_*}_{III}(z=6) \sim 10^{-7} \; \rm{M_\odot yr^{-1} Mpc^{-3}}$, or a Population III star formation rate of 0.001 per cent of the total at $z=6$.  These star formation rates are similar to rates predicted from pockets of metal free gas that may exist at low redshift \citep{tornatore/etal:2007, trenti/etal:2009}. 

\begin{figure}
\centering \noindent
\includegraphics[width=8.8cm]{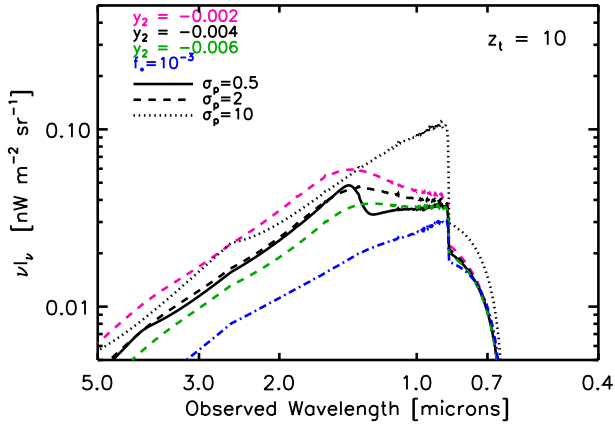}
\caption{The effect on the spectrum of the NIRB from stars above a redshift of 6 by changing $\sigma_p$, which describes the width of the transition from Population III to Population II stars.  A larger $\sigma_p$ allows this transition to be more extended in time.  Also shown is the dependence on the star formation rate, described by the parameter $y_2$, which affects the steepness of the star formation rate as a function of z.  Finally, we look at the case where the fraction of baryons ($f_*$) that form into stars over time is constant.}
\label{fig:sigfesc}
\end{figure}

The transition from Population III to Population II stars is also described by the time that it takes this transition to occur, parametrized by $\sigma_p$.  In Fig. \ref{fig:sigfesc}, $\sigma_p$ is adjusted, so as to cause the transition to be abrupt ($\sigma_p=0.5$) or extended in time ($\sigma_p=10$).  When $\sigma_p=10$, there is still a considerable amount of Population III stars forming at $z\sim 6$.  Because of this, there is significant Lyman-$\alpha$ emission at shorter wavelengths from these massive stars.  If $\sigma_p$ is smaller, this low-redshift contribution from massive Population III Lyman-$\alpha$ emission falls, causing a drastic change in shape of the Lyman-$\alpha$ bump.  If the transition is very abrupt, there can actually be more emission from more distant Population III stars than nearby Population II stars, creating a situation where there is more emission at longer wavelengths than shorter wavelengths.  Therefore, changing $\sigma_p$ can change the shape of the Lyman$-\alpha$ bump.

Also shown in Fig. \ref{fig:sigfesc} is the effect of changing the slope of the star formation rate as a function of redshift, described by our free parameter $y_2$ (in equation \ref{eq:sfrapp}), where more negative values of $y_2$ lead to a more rapidly decreasing value of the star formation rate.  Changing this free parameter has only a slight effect on the spectrum of the NIRB.  This is because changing $y_2$ has the greatest effect at high redshift (since equation \ref{eq:sfrapp} is normalized at $z\sim 6$).  The stars at higher redshift, since they are more distant, have a smaller contribution on the overall shape of the spectrum of the NIRB.  

However, instead of assuming a star formation rate as a function of redshift via equation \ref{eq:sfrapp}, we can instead assume the fraction of baryons that form into stars ($f_*$) is constant with redshift.  (This method was used in previous papers, such as \citet{fernandez/etal:2010}, \citet{fernandez/etal:2012fluc}, and \citet{cooray/etal:2012}).  The star formation rate equivalent to this assumption is shown by the pink dotted line in Fig. \ref{fig:varsfra}, for $f_* = 0.003$\footnote{This value was chosen so that the star formation rate is approximately consistent with our other models at $z \sim 6$.}, derived by combining $f_*$ with the collapse fraction of haloes, and assuming a time formation time-scale (here, 11.5 Myr, like in \citet{fernandez/etal:2012fluc, iliev/etal:2012}).  The star formation rate as a function of redshift for this case is much steeper.  Because of the lower level of the star formation rate at higher redshift, the Lyman-$\alpha$ bump is not present, due to lower redshift Population II stars dominating over any Population III contribution.  The spectrum for this case is shown as the blue dashed-dotted line in Fig. \ref{fig:sigfesc}.  It is important to note that the assumption of a constant $f_*$, as often assumed in the literature, along with a constant star formation time-scale, predicts a lower star formation rate at high redshift in relation to lower redshifts as compared to our other models.  This would erase any spectral signature of the Lyman-alpha bump.

\subsection{The Masses of Population III Stars}

\begin{figure}
\centering \noindent
\includegraphics[width=8.8cm]{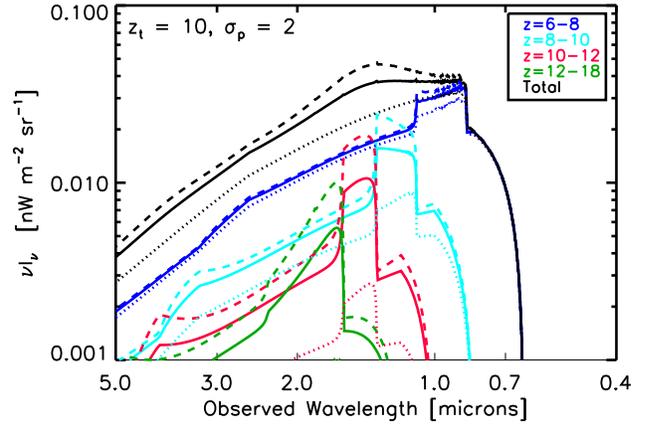}
\caption{The spectrum of the NIRB when Population III stars have various stellar masses.  In each case, the Population II stars have the same mass spectrum.  These are combined with Population III stars with a heavy Larson mass spectrum (with $m_c=250 \; \rm{M_\odot}$ - dashed lines), Population III stars with a light Larson mass spectrum (with $m_c=10 \; \rm{M_\odot}$ - solid lines), and Population III stars with a Salpeter mass spectrum (dotted lines).}  
\label{fig:mass}
\end{figure}

To model the possibility that Population III stars might not be as massive, we computed the spectrum of the NIRB when the value of $m_c$ for Population III stars is $10\; \rm{M_\odot}$ and the mass limits of the mass spectrum are $m_1 = 0.1 \; \rm{M_\odot}$ and $m_2=150\; \rm{M_\odot}$ (see equation \ref{eq:larson}).  In addition, we also calculate the spectrum of the NIRB if Population III stars have a Salpeter slope with $m_1 = 0.1 \; \rm{M_\odot}$ and $m_2=150\; \rm{M_\odot}$.  The results of this test, in comparison to our reference case with heavier Population III stars, are shown in Fig. \ref{fig:mass}.  

When Population III stars are more massive, there is a more pronounced bump from the Lyman-$\alpha$ line from the end of the Population III era.  But, if Population III stars are no more massive than Population II stars, the spectrum is smooth, and the Lyman-$\alpha$ bump is not seen at all.  More massive Population III stars would emit a stronger Lyman-$\alpha$ emission line, therefore the presence of a Lyman-$\alpha$ bump would be the hallmark of a massive Population III era.  

\subsection{The Contribution of the IGM and the Escape Fraction}
\label{sec:IGM}
Nebular emission will originate from within the halo, or, if some photons escape the halo, from the IGM.  However, because of the low density of the IGM, the IGM component of the NIRB might be quite low.  Therefore, it is possible that the contribution of the IGM to the overall spectrum of the NIRB can be negligible \citep{ nakamoto/etal:2001, cooray/etal:2012}.  

Thus far, we have assumed that the IGM does not contribute at all to the overall intensity of the NIRB.  In this case, nebular emission will only originate from the haloes themselves, so their luminosity must be multiplied by the factor of $(1 - f_{esc})$. 
If, on the other hand, the recombination time within the IGM is very short, the IGM will contribute to the spectrum of the NIRB.  
However, the spectrum of the NIRB is very weakly dependent on whether or not the IGM is contributing to the overall emission.  This is because, for this case, the escape fraction of photons is only $0.16$, set to be consistent with constraints from reionization.  Therefore, the nebular component of the emission is only diminished by $\sim 16$ per cent when the IGM is not contributing at all to the emission in the NIRB.  This is not enough to make a significant change in the spectrum of the NIRB.

Yet, changing the value of the escape fraction can affect the spectrum of the NIRB.  \citet{yue/etal:2012} did not see a Lyman-$\alpha$ bump in their predicted spectrum of the NIRB. This was because they allowed $f_{esc}$ to vary with redshift.  According to their model, $f_{esc}(z=5)=0.05$, and rises with redshift, reaching unity at $z\sim11$.  Because of this, any Lyman-$\alpha$ contribution from stars at higher redshift would be suppressed, and therefore, no Lyman-$\alpha$ bump from these stars would be seen.  Therefore, the presence of a Lyman-$\alpha$ bump would also indicate that $f_{esc}$ is low enough to allow a prominent Lyman-$\alpha$ feature at higher redshifts.

It is also possible that the escape fraction for Population III stars is above that of Population II stars (especially if the Population III stars are significantly massive).  This case would further smooth out any effect of the Lyman-$\alpha$ bump, since the Lyman-$\alpha$ emission from higher redshift Population III stars in haloes would be diminished.


\section{Conclusions}
\label{sec:conclusions}

The spectrum of the NIRB can give information on the end of the Population III era.  This is mainly reflected in the presence and shape of the 'Lyman-$\alpha$ bump', whose shape reflects not only when Population III stars transitioned to Population II stars, but also properties of these stars as well.  
The presence of this bump is mostly dictated by the relative presence of massive Population III stars to less massive Population II stars.  A larger bump corresponds to a situation where more massive Population III stars exist to lower redshifts.  For our reference case, this bump in the spectrum is seen when $z_t \lesssim 15$ (or, in other words, a case where the Population III star formation rate is greater than $\sim 10^{-10}$ times that of the total star formation rate).    

The shape of the Lyman-$\alpha$ bump also will reveal information on how quickly the Population III era ended.  An extended tail of Population III to low redshifts, which would result from a very long transition from Population III to Population II stars, would result in a strong Lyman-$\alpha$ peak.  However, if the transition is abrupt, the Lyman-$\alpha$ bump will change shape, such that it is possible that there is more intensity at longer wavelengths than shorter wavelengths, a result of more emission from more massive, more distant Population III stars than nearer, less massive Population II stars.  In addition, the wavelength at which the spectrum peaks at will be indicative of when massive Population III stars existed.    

However, if no Lyman-$\alpha$ bump is seen at all, this indicates that the Population III era ended much earlier, with $z_t \gtrsim 15$, or that Population III stars are not very massive, being closer to mass to the mass spectrum we see today.  On the other hand, it can also indicate that the escape fraction increases at high redshifts so that the Lyman-$\alpha$ emission from higher redshift Population III stars is suppressed, and that the emission from the IGM is negligible.    

It may be possible to observe this spectral feature, independent of complete subtraction of foregrounds.  Even though this observation is very difficult to perform, the relative change in intensity resulting from a Lyman-$\alpha$ bump as a function of wavelength would be a unique spectral signature, different from the foregrounds of both $z<6$ galaxies and Zodiacal light.  However, it may be difficult to understand if the presence of a Lyman-$\alpha$ bump is due to a late transition from Population III to Population II stars (parametrized by $z_t$) or by another effect, such as a rapidly changing star formation rate as a function of redshift or evolution of the escape fraction.  Yet regardless of the cause, if a bump is seen, it is indicative that a non-negligible amount of massive Population III stars exist even to late times ($z\lesssim 8 -12$).   

These results have interesting consequences.  It is possible that small pockets of metal-free gas persist until low redshifts, so that Population III star formation might still persist at $z \sim 6$  \citep{tornatore/etal:2007, trenti/etal:2009}.  Even small Population III star formation rates ($\dot{\rho_*}(z=6) \sim 4 \times 10^{-5}$ or $2 \times 10^{-7} \; \rm{M_\odot yr^{-1} Mpc^{-3}}$ for our cases $\sigma_t=2$ and $z_t = 10$ and $12$ respectively), could actually lead to a signature on the spectrum of the NIRB.  Of course, if these pockets of metal free gas are indeed rare on the sky, measurements must be carefully done to assure this effect is observed.

\section{Acknowledgments}
We would like to thank Michele Trenti for helpful conversations.  We also acknowledge  the Dutch Science organization (NWO) Vernieuwingsimpuls Vici program for support.


 \newcommand{\noop}[1]{}

\end{document}